\documentclass[review,12pt]{elsarticle}

\usepackage{color}
\usepackage{amssymb}
\usepackage{amsmath}
\usepackage{setspace}
\usepackage{graphicx}
\usepackage{subfigure}
\usepackage{booktabs}
\usepackage{makecell}
\usepackage{adjustbox}
\usepackage[ruled,linesnumbered]{algorithm2e}
\journal{Computer Methods in Applied Mechanics and Engineering}

\begin{document}
	
	\begin{frontmatter}
		
		
		
		\title{Dynamical pressure boundary condition for weakly-compressible 
			smoothed particle hydrodynamics}
		
		
		\author[a]{Shuoguo Zhang}
		\author[a]{Yu Fan}
		\author[a]{Dong Wu}
		\author[a]{Chi Zhang}
		\author[a]{Xiangyu Hu\corref{cor1}}
		\ead{xiangyu.hu@tum.de}
		
		\cortext[cor1]{Corresponding author}				
		\address[a]{School of Engineering and Design, 
			Technical University of Munich, Garching, 85748, Germany}
		%%\affiliation{organization={},%Department and Organization
		
		\begin{abstract}		
This paper introduces a novel dynamical pressure boundary condition 
for weakly-compressible smoothed particle hydrodynamics (WCSPH). 
Unlike previous methods that rely on indirect approaches or ghost particles, 
our method integrates the dynamical boundary pressure directly into 
the SPH approximation of the pressure gradient on near-boundary particles. 
Additionally, we develop a meshfree bidirectional in-/outflow buffer by 
periodically relabelling buffer particles at each time step, a concept that 
has not been explored before. 
This simple yet effective buffer facilitates the simulation of 
both uni- and bidirectional flows, 
especially those with mixed in-/outflow boundary conditions. 
We validate the accuracy and convergence of our method through benchmark cases 
with available analytical solutions. 
Furthermore, we demonstrate its versatility in hemodynamic simulations 
by investigating generic carotid and aorta flows with the Windkessel model, 
paving the way for studying the cardiovascular system 
within a unified meshfree computational framework.
		\end{abstract}
		\begin{graphicalabstract}
		\end{graphicalabstract}
		
		\begin{highlights}
			\item A dynamical pressure boundary condition for 
			the WCSPH method to impose constant or time-dependent pressure.  
			\item A much simpler bidirectional in-/outflow buffer 
			for the SPH method. 
			\item Preliminary hemodynamic simulations on carotid 
			and aorta flows with the Windkessel model.
		\end{highlights}
		
		
	\end{frontmatter}
	
	
	\section{Introduction}
	\label{Introduction}
	The hemodynamic flows, 
	as typical incompressible fluid dynamics applications in bio-engineering, 
	are very often defined by multiple dynamical or time-dependent in-/outflow
	and mix-in-/outflow 
	(coexistence of local forward and reverse flows at a single in-/outlet surface) 
	boundaries with complex geometrics \cite{Catanho2012Model}.
	Since determining the cross-section velocity profile at these boundaries is 
	much more difficult than measuring the boundary pressure 
	as the latter is merely time-dependent and can be considered as constant 
	along the entire cross-section surface, 
	imposing pressure, other than velocity, 
	boundary conditions is often preferred in hemodynamic flow simulations.
	For grid-based methods, 
	despite the implementation of dynamical pressure boundary conditions 
	is well-established, 
	its employment for practical hemodynamic problems is often very difficult
	because of the intrinsic limitation	on modeling 
	the associated complex fluid-structure interaction (FSI), 
	especially when mesh topological changes are involved.
	On the other hand, although the meshfree SPH (smoothed particle hydrodynamics) 
	method is free of mesh topologies and 
	able to model complex FSI problems effectively under 
	a unified computational framework 
	\cite{antoci2007numerical, han2018sph, Zhang2021CPC}, 
	it has not been widely employed for hemodynamic flow simulations due to 
	the remaining well-known challenge of pressure boundary conditions.
	
	In the most popular weakly compressible SPH (WCSPH) method, 
	since the velocity field and pressure are not directly coupled due to 
	the approximation of incompressible condition, 
	a straightforward approach by which 
	the particle velocity at the boundary can be updated from 
	given pressure condition directly is yet to be developed.
	Holmes et al. \cite{HOLMES2021Novel} obtains the boundary particle velocity 
	from a given constant pressure at simple planer in-/outflow boundaries 
	indirectly through an elaborate correction
	involving the continuity equation and the artificial equation of state. 
	It is still unclear whether this approach can be effectively employed for 
	practical cases with dynamical pressure, mix-in-/outflow boundaries of complex geometries.
		
	While the decoupling between particle pressure and velocity 
	is not the case for incompressible SPH (ISPH) method, 
	the implementation for general pressure boundary condition
	is still troublesome due to the ineffectiveness of particle handling,
	such as	populating ghost particles, generating and deleting fluid particles 
	at in-/outlets.
	For example, Hirschler et al. \cite{HIRSCHLER2016Open} 
	and Kunz et al. \cite{KUNZ2016Inflow} employed 
	a mirror-axes particle technique to generate ghost and fluid particles 
	for specifying constant pressure at simple domain boundaries.
	However, no matter using fixed or moving axes, 
	significant numerical errors characterized by the artifacts of 
	void regions and disorder particles are produced at the in-/outlets. 
	In comparison, 
	Monteleone et al. \cite{Alessandra2017} developed a grid-based approach for generating 
	ghost particles and a very complex geometric method using conical scan 
	to generate inflow fluid particles 
	at the in-/outlets. Together with further elaborative modifications on the coefficient 
	matrix and the right-hand-side term of the pressure Poisson equation, the method is able 
	to avoid the artifacts produced in Ref. \cite{HIRSCHLER2016Open, KUNZ2016Inflow}.
	
	In this paper, we develop a dynamical pressure boundary condition for WCSPH 
	to simulate hemodynamic flow with in-/outlets with general complex geometries.
	With the assumption of zeroth-order consistency in SPH discretization,
	the present dynamic boundary pressure is imposed directly 
	to compute the near-boundary gradient 
	and to circumvent the cumbersome handling of ghost particles 
	as in previous work. Different from Ref. \cite{HOLMES2021Novel}, 
	the present particle velocity at the boundary is obtained straightforwardly 
	from the discretized momentum conservation equation.
	Also different from Ref. \cite{Alessandra2017}, 
	the present meshfree method does not rely on grid-based approach 
	but a bidirectional buffer for particle handling, 
	i.e. generating and deleting fluid particles at the in-/outlets.
	Although the present particle handling is quite simple,
	it does not produce artifacts as in Refs. \cite{HIRSCHLER2016Open, KUNZ2016Inflow}.
	Note that, the buffer used in this paper is relevant to those 
	in several other SPH boundary conditions, 
	such as periodic \cite{Braun2015, Molecular2019, Negi2022how}, 
	open \cite{Martin2009, FEDERICO2012Simula,
		Vacondio2012mod, Alvarado-Rodriguez2017, Ferrand2017, Tafuni2018, Negi2020} 
	and free-stream \cite{Shuoguo2022free} conditions.
	The difference is that the present buffer, at the first time, 
	is able to cope with the mix-in-/outflow, where particle generation and deletion 
	are carried out at a single boundary surface,
	helped by periodically relabelling buffer particles at each time step. 

	The remainder of this paper is organized as follows. 
	First, Section \ref{Riemann-based_WCSPH_method} gives a brief overview of 
	the underlying WCSPH method.
	In Section \ref{Pressure_boundary_condition}, 
	the proposed dynamical pressure boundary condition is detailed. 
	The accuracy, convergence and applicability of the developed method are 
	demonstrated by 
	several flow examples in Section \ref{Testing_and_verification}, 
	including the startup of Poiseuille and Hagen-Poiseuille flows, 
	PIVO (Pressurized Inlet, Velocity Outlet) and VIPO 
	(Velocity Inlet, Pressure Outlet) channel flows,
	pulsatile channel flow, and generic carotid and aorta flows 
	coupling with Windkessel model for the dynamical boundary pressure.
	Finally, concluding remarks are given in Section \ref{Conclusion}. 
	The code accompanying this work is implemented in 
	the open-source SPHinXsys library
	\cite{Zhang2021CPC}, 
	and is available at the project website https://www.sphinxsys.org 
	and the corresponding Github repository.
	\section{Weakly-compressible SPH method}
	\label{Riemann-based_WCSPH_method}
	\subsection{Governing equations}
	\label{Governing_equations}
	Within the Lagrangian framework, the mass- and momentum-conservation equations 
	are respectively written as 

	\begin{equation} \label{continuity_equation}		
	 \frac{d\rho}{dt} = -\rho\nabla\cdot\mathbf{v},
	\end{equation}
    and

    \begin{equation} \label{momentum_conservation_equation}	
	\frac{d\mathbf{v}}{dt} = -\frac{1}{\rho}\nabla p+\nu\nabla^{2}\mathbf{v},
    \end{equation}
	where $ \rho $ is the density, $t $ the time, $ \mathbf{v} $ the velocity, $ p $ 
	the pressure, and $ \nu $ the kinematic viscosity. Under the weakly-compressible 
	assumption, 
	the system of Eqs.(\ref{continuity_equation}) and (\ref{momentum_conservation_equation}) 
	are closed by the artificial isothermal equation of state (EoS) 
	\begin{equation} \label{state_equation}
		p=c_0^2(\rho-\rho_{0}), 
	\end{equation}
    where $c_{0}=\lambda U_{max}$ represents the artificial sound speed with $U_{max}$ 
    indicating 
    the maximum anticipated flow speed, and $\rho_{0}$ the initial reference
    density. In WCSPH method, sufficiently large coefficient $ \lambda $ 
    is selected to ensure near incompressible behavior 
    \cite{HOLMES2021Novel, Monaghan1994Sim, Morris1997Modeling}.
	
	\subsection{SPH discretization}
	\label{SPH_discretization}
	Following Refs.\cite{Zhang2021CPC, Hu2006multi, Zhang2017trans, Zhang2020Dual, ZHANG2021An}, 
	both the mass- and momentum-conservation equations Eqs.(\ref{continuity_equation}) 
	and (\ref{momentum_conservation_equation}) are discretized using the Riemann-based SPH 
	scheme, in respect to particle $i$

	\begin{equation} \label{disctretized_continuity_equation}
		\frac{d\rho_{i}}{dt}=2\rho_{i}\sum_{j}\frac{m_{j}}{\rho_{j}}(\mathbf{v}_{i}-
		\mathbf{v}^{*}_{ij})
		\cdot\nabla W_{ij}, 
	\end{equation}
    and

	\begin{equation} \label{disctretized_momentum_conservation_equation}
	\frac{d\mathbf{v}_{i}}{dt}=-2\sum_{j}m_{j}(\dfrac{P^{*}_{ij}}{\rho_{i}\rho_{j}})\nabla 
	W_{ij}+2\sum_{j}m_{j}\frac{\eta\mathbf{v}_{ij} }{\rho_{i}\rho_{j}r_{ij}}\frac{\partial W_{ij}}
	{\partial r_{ij}},
    \end{equation}
	where $ m $ is the particle mass, $ \eta $ the dynamic viscosity, subscript $ j $ the 
	neighbor
	particles, and $\mathbf{v}_{ij}=\mathbf{v}_{i}-\mathbf{v}_{j} $ the relative velocity 
	between 
	particles $ i $ and $ j $. Also, $ \nabla W_{ij} $ denotes the gradient of the kernel 
	function $ W(|\mathbf{r}_{ij}|,h) $, where  $ \mathbf{r}_{ij}=\mathbf{r}_{i}-\mathbf{r}_{j} $, 
	and $ h $ the smooth length. 
	Furthermore, $ \mathbf{v}^{*}_{ij} = U^{*}_{ij}\mathbf{e}_{ij}+(\mathbf{\overline{v}}_{ij}-\overline{U}_{ij}
	\mathbf{e}_{ij})$ with $ \mathbf{e}_{ij}=\mathbf{r}_{ij}/r_{ij} $, 
	$ \mathbf{\overline{v}}_{ij} =(\mathbf{v}_{i}+\mathbf{v}_{j})/2 $ 
	the average velocity between particles $ i $ and $ j $ 
	and $\overline{U}_{ij} =\mathbf{\overline{v}}_{ij} \cdot \mathbf{e}_{ij}$. Here, 
	$P^{*}_{ij}$ and  $ U^{*}_{ij}$ are obtained by solving the inter-particle one-dimensional 
	Riemann problem with a low dissipation limiter as given by Ref. \cite{article2017Chi}. 
    
    In order to increase the computational efficiency, 
the dual-criteria time stepping method \cite{Zhang2020Dual} is employed here.
Therefore, the particle cell-linked lists and neigbor configurations are only rebuilt for 
each advection time step (referred to as time step herein) 
and kept unchanged for the several acoustic time-steps.
One can refer Ref.  \cite{Zhang2020Dual}  details.
In addition, to decrease the accumulated density error during long-term simulations, 
    a density initialization scheme \cite{Zhang2021CPC} is also used to 
    stabilize the density field prior the acoustic time steps as
	\begin{equation} \label{density_summation}
		\rho_{i}=\rho_{0}\dfrac{\sum W_{ij}}{\sum W^{0}_{ij}},
	\end{equation}
    where the superscript $ 0 $ represents the reference value in the initial configuration.
    \subsection{Transport-velocity formulation}
    \label{Transport_velocity_formulation}
    To avoid particle clumping and void regions in SPH simulations where the tensile instability 
    is present \cite{Lind2012Incom, OGER201676}, regularization is implemented 
    to maintain a uniform particle distribution \cite{litvinov2015towards, Negi2022Tec}.
    Currently, the particle shifting technique (PST) 
    \cite{Lind2012Incom, OGER201676, Negi2022Tec, SKILLEN2013163Incom, KHAYYER2017236Comp, SUN201725} 
    and the transport-velocity formulation 
    (TVF) \cite{Zhang2017trans, Adami2013trans} are two typical schemes to address this issue. 
    In the present work, the TVF scheme is adopted, and the particle advection velocity 
    $ \widetilde{\mathbf{v}} $ is rewritten as

    \begin{equation} \label{transport_velocity}
    	\widetilde{\mathbf{v}}_{i}(t + \delta t)=\mathbf{v}_{i}(t)+
    	\delta t\left(\frac{\widetilde{d}\mathbf{v}_{i}}{dt}-p_{max}\sum_{j} \frac{2m_{j}}
    	{\rho_{i}\rho_{j}} \frac{\partial W_{ij}}{\partial r_{ij}}\mathbf{e_{ij}} \right),
    \end{equation}
    where the global background pressure $p_{max}=7\rho_{0}\mathbf{v}_{max}^{2}$ 
    with $ \mathbf{v}_{max} $ denoting the maximum particle velocity 
    at each time step \cite{Zhang2020Dual}.
    
    Since the TVF scheme is applicable only to inner fluid particles far away from in-/outlet 
    boundaries \cite{Adami2013trans}, accurately identifying boundary particles 
    is a necessary prior. To avoid misidentifying and missing 
    boundary particles \cite{Shuoguo2022free, Dilts, Haque, Lee2008}, 
    the spatio-temporal identification approach \cite{Shuoguo2022free} is utilized in 
    present work, where approximately three layers of fluid particles are identified 
    as boundary particles.    
    	
	\section{Dynamical pressure boundary condition}
	\label{Pressure_boundary_condition}	
	\subsection{Bidirectional in-/outflow buffer}
	\label{Bidirectional_buffer}
	\begin{figure}[htb]
		\centering     
		\includegraphics[width=0.9\textwidth]{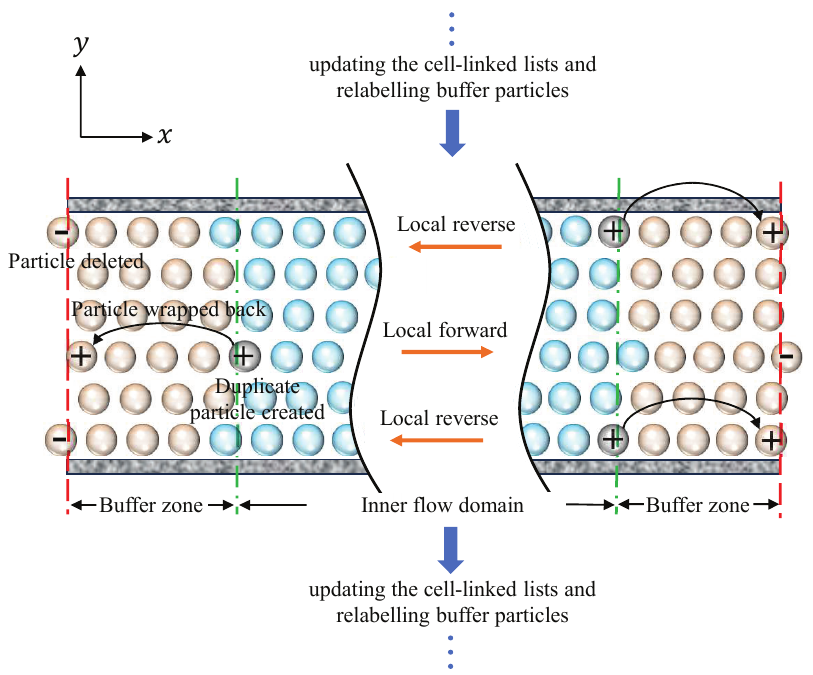}
		\caption{The illustration of particle handling for the bidirectional in-/outflow buffer. 
		Within the left and right buffers, the dashed red lines indicate the in-/outlet bound, 
		and the dot-dashed green lines connect the buffers to the inner flow domain.  
		The inner-domain, buffer and duplicate particles are colored with blue, gold and gray, 
		respectively. Note that, the wall end should be aligned to the buffer surface.}
		\label{particle_deletion_generation}
	\end{figure}

	In Figure \ref{particle_deletion_generation}, 
	a two-dimensional bidirectional 
	flow along $x$-axis between two parallel walls is sketched 
	to demonstrate the particle handling
	of the bidirectional in-/outflow buffer. 
	To realize the bidirectional buffer, the key step is to relabel buffer particles 
	(indicating the fluid particles located within the buffer zone) at the end of each 
	time step. For example, buffer and inner-domain particles are assigned the integer labels 1 and 0, 
	respectively. 
	Note that, as the present smoothing length $ h = 1.3dp $,
	with $ dp $ denoting the initial particle spacing, is used, 
	the buffer should consist of at 
	least three layers of particles to ensure full support for 
	the inner-domain particles next to the buffer.
	
	Here, since the right buffer operates based on the same principle, 
	only the left buffer in Figure \ref{particle_deletion_generation} is used to 
	detail the present method. 
	For a local forward flow at the present time step, as shown in Figure 
	\ref{particle_deletion_generation},
	the buffer will generate new particles. 
	In this case, some buffer particles cross the right buffer bound. 
	Duplicate particles are generated with the same states as the crossing ones, 
	and then treated as ordinary particles of the inner flow domain. 
	Meanwhile, the original buffer particles will be recycled from the left buffer 
	bound at the periodic positions \cite{Tafuni2018, Shuoguo2022free}. 
	Correspondingly, if the buffer particles leave the left buffer bound in a local 
	reverse flow, as shown in Figure \ref{particle_deletion_generation}, they will 
	be deleted as outflow particles.
	Since some fluid particles from inner domain may enter the buffer zone from the 
	right buffer bound at the present time step, all fluid particles within the buffer 
	zone will be relabeled at the end of present time step (after the cell-linked 
	lists have been rebuilt). 
	Note that, this can be easily achieved by checking whether the particles in the 
	nearby updated cell-linked lists are located within the buffer zone. 

\begin{figure}[htb]
		\centering     
		\includegraphics[width=0.79\textwidth]{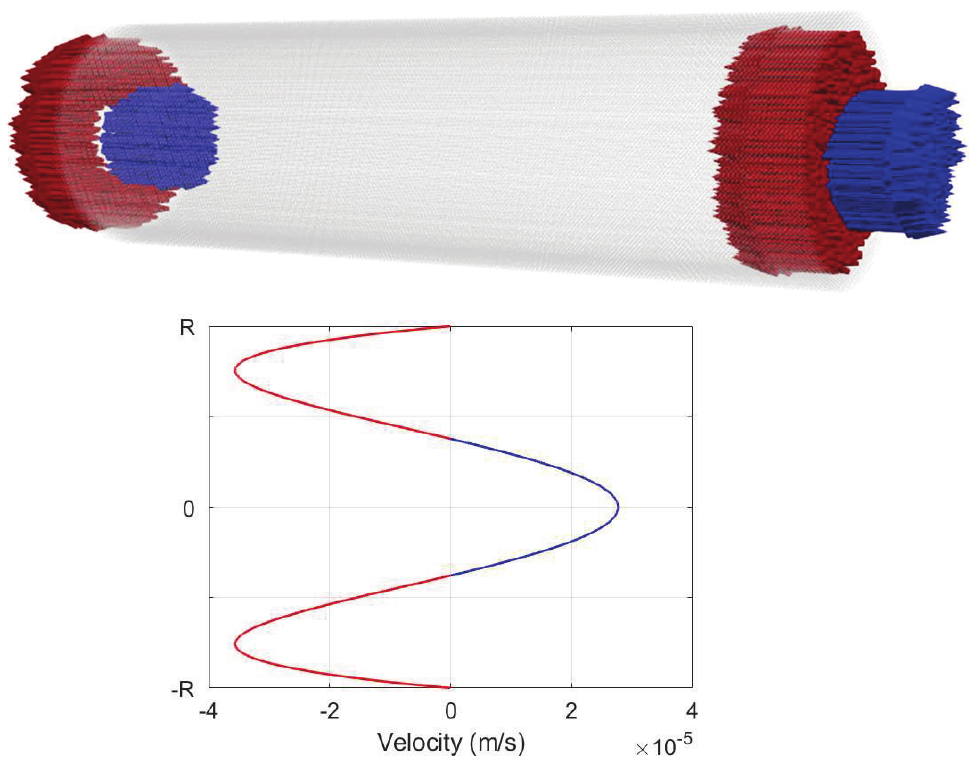}
		\caption{Velocity profile with mixed forward and reverse flows at the in-/outlet 
			boundary (top panel), corresponding to the analytical velocity profile 
			(bottom panel) 
			at $ t=7.975s $ of the pulsatile flow in Subsection \ref{Pulsatile_flow}. 
			The velocity 
			vector in top panel is scaled to a uniform length, 
			with the flow direction indicated by the arrow.}
		\label{bidirectional_flow}
	\end{figure}

	It is worthy to emphasize that the present bidirectional buffer copes with the 
	mix-in-/outflow boundary in a natural way.
	At the right buffer bound, the inflow particles are created and the outflow 
	particles are identified as new buffer particles at the end of time step.
	At the left buffer bound, while the inflow is handled by recycling the buffer 
	particles leaving the right buffer bound, the outflow particles are simply deleted.  
	Figure \ref{bidirectional_flow} gives typical mix-in-/outflow profile obtained 
	by the present method.

	\subsection{Imposing dynamical boundary pressure}
	\label{Boundary_target_pressure}
	\begin{figure}[htb]
		\centering     
		\includegraphics[width=0.57\textwidth]{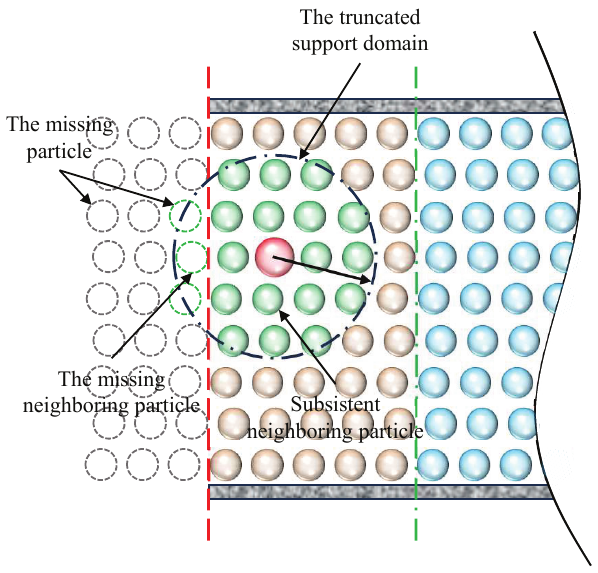}
		\caption{The truncated support domain of fluid particles near the in-/outlet.}
		\label{implementating_targer_pressure}
	\end{figure}
    As shown in Figure \ref{implementating_targer_pressure}, the support domain of fluid 
    particles near the in-/outlet (also referred to as near-boundary particles herein) is 
    truncated, with the postulated "missing" particles beyond the buffer bound.
    For pressure boundary condition,
    the given boundary pressure $ p_{target} $ needs to be specified 
    as the evolution of the in-/outflow velocity 
    is strongly influenced by the corresponding near-boundary pressure gradient.  
     
    For a particle with full kernel support, 
    the standard SPH discretization of gradient 
    \cite{Monaghan1992Smo} approximates the zeroth-order consistent, 
    i.e., $ \sum_{j}\frac{m_{j}}{\rho_{j}}\nabla W_{ij} \approx \mathbf{0}$. 
    In this paper, this property is utilized to 
    approximate pressure gradient for a near-boundary particle. 
    Specifically, a near-boundary particle $i$ has 
    \begin{equation} \label{zeroth-order consistent}
	\sum_{j}\frac{m_{j}}{\rho_{j}}\nabla W_{ij}+\sum_{k}\frac{m_{k}}{\rho_{k}}
	\nabla W_{ik}  \approx  \mathbf{0}, 
    \end{equation}
    with $ j $ and $ k $ represent the subsistent and "missing"
    neighboring particles, respectively.
    Referring Eq. (\ref{disctretized_momentum_conservation_equation}),
    the SPH approximation of the pressure gradient 
    at a near-boundary particle can be written as
    \begin{equation} \label{pressure_gradient_component}
    	\nabla p_{i}  =2 
    	\sum_{j}P^{*}_{ij}\frac{m_{j}}{\rho_{j}}\nabla W_{ij}+2\sum_{k}P^{*}_{ik}
    	\frac{m_{k}}{\rho_{k}}\nabla W_{ik}.
    \end{equation}
    By using Eq.(\ref{zeroth-order consistent}) 
    and assuming uniform in-/outlet pressure as $ p_{target} = P^{*}_{ik}$,
    Eq. (\ref{pressure_gradient_component}) can be 
    rewritten as 
    \begin{equation} \label{rewritten_pressure_gradient_component}
	\nabla p_{i}  =2 
	\sum_{j}P^{*}_{ij}\frac{m_{j}}{\rho_{j}}\nabla W_{ij}-2p_{target}\sum_{j}\frac{m_{j}}
	{\rho_{j}}\nabla W_{ij}.
    \end{equation}
    Then, the modified Eq.  (\ref{disctretized_momentum_conservation_equation}) 
    for near-boundary particles has the form 
    \begin{equation} \label{modified_discretized_momentum_equation}
    	\begin{split}
    	\frac{d\mathbf{v}_{i}}{dt}=&-2\sum_{j}m_{j}(\dfrac{P^{*}_{ij}}{\rho_{i}
    		\rho_{j}})\nabla W_{ij}+2p_{target}\sum_{j}(\dfrac{m_{j}}{\rho_{i}
    		\rho_{j}})\nabla W_{ij}  
    	\\&+2\sum_{j}m_{j}\frac{\eta\mathbf{v}_{ij} }{\rho_{i}\rho_{j}r_{ij}}
    	\frac{\partial W_{ij}}{\partial r_{ij}}
        \end{split}.
    \end{equation}
    Note that, the present formulation does not require ghost particles 
    to replace the postulate "missing" particle as in previous work. Furthermore, 
    as the second term on the right side of Eq. (\ref{modified_discretized_momentum_equation}) 
    vanishes for particles with full kernel support, Eq. (\ref{modified_discretized_momentum_equation}) 
    is applied to all buffer particles in present work, without specifically identifying 
    near-boundary particles with truncated kernel support.
	Also note that, the velocity updated by Eq.(\ref{modified_discretized_momentum_equation}) 
    should be perpendicular to the in-/outlet boundary, 
    aligning with the in-/outflow condition \cite{HOLMES2021Novel}
    \begin{equation} \label{velocity_correction}
    	\mathbf{v}_{i}=(\mathbf{v}_{i}\cdot \hat{u})\hat{u}, 
    \end{equation}
    where $ \hat{u} $ is the unit normal vector of 
    the in-/outlet boundary or buffer bound. 
    
    To cope with the truncation for density summation, 
    the reinitialization of Eq. (\ref{density_summation}) 
    is not carried out for near-boundary particles (i.e., buffer particles herein),
    and the density of the newly populated (actually recycled) particles 
    in the bidirectional in-/outflow buffer is obtained
    following the boundary pressure and EoS as
    \begin{equation} \label{postulate-density}
	\rho_{i} = \rho_{0} + p_{target} / c_0^2.
    \end{equation}

    When the bidirectional in-/outlet buffer works with the velocity in-/outflow 
    boundary condition, such as in PIVO (Pressurized Inlet, Velocity Outlet) and VIPO 
    (Velocity Inlet, Pressure Outlet) flows, the pressure boundary 
    condition should also be imposed at the velocity in-/outlet to 
    eliminate the truncated error in approximating pressure gradient, 
    but the corresponding $ p_{target} $ in Eq.(\ref{modified_discretized_momentum_equation}) 
    is given as $ p_{i} $. Meanwhile, both the density
    and pressure of newly populated particles remain unchanged.
    
   	\section{Numerical examples}
	\label{Testing_and_verification}
	In this section, to validate the present method, a set of benchmark 
	cases are simulated and the results are compared with those of previous 
	experiments and simulations. 
	These cases including the startup of Poiseuille and Hagen-Poiseuille 
	flows, PIVO  and VIPO  channel flows, and carotid and aorta flows 
	coupling with Windkessel model for the dynamical boundary pressure. 
	The 5th-order Wendland kernel \cite{Wendland1995} is
	employed in all the following simulations. 
	As for the treatment of wall boundary, 
	a one-sided Riemann solver is employed and we refer to Ref. \cite{article2017Chi}
	for more details.
	\subsection{Startup of Poiseuille flow}
	\label{Poiseuille_flow}
\begin{figure}[htbp]
		\centering     
		\includegraphics[width=0.8\textwidth]{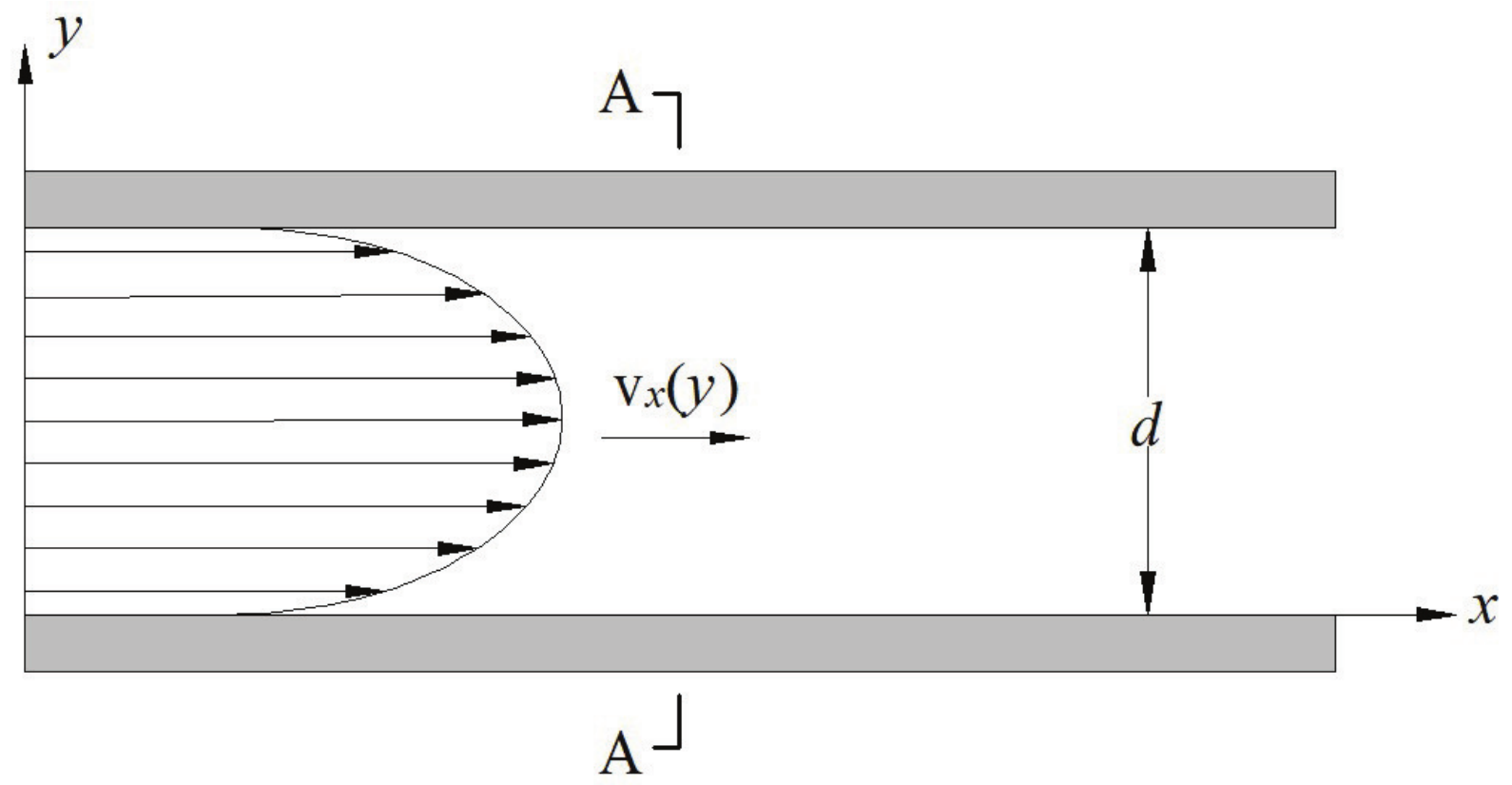}
		\caption{Schematic of Poiseuille flow. Velocity profile is extracted 
			at the cross-section A in the simulation.}
		\label{Poiseuille_flow_schematic}
	\end{figure}

	As a well-defined benchmark to verify the pressure boundary condition, 
	the startup of two-dimensional Poiseuille flow 
	\cite{HOLMES2021Novel, Morris1997Modeling, Takeda1994Nume, Sigalotti2003SPH, Holmes2011Smoo} 
	is studied herein. 
	The analytical velocity evolution $ \mathbf{v}_{x}(y,t) $ is given as

	\begin{equation} \label{Poiseuille_profile}
		\begin{split}
			&\mathbf{v}_{x}(y,t) = \frac{\mathtt{\Delta} P}{2\eta L}y(y-d) \\	
			&+ \sum_{n=0}^\infty \dfrac{4\mathtt{\Delta} Pd^{2}}{\eta L\pi^{3}
				(2n+1)^{3}}sin(\dfrac{\pi y}{d}(2n\\&+1))
			\exp(-\dfrac{(2n+1)^{2}\pi^{2}\eta}{\rho d^{2}}t)
		\end{split},
	\end{equation}
	where $ y \in (0,d)$ with $ d $ denoting the distance between plates, and 
	$ \mathtt{\Delta} P $ the pressure drop across a flow length of $ L $.
	Figure \ref{Poiseuille_flow_schematic} shows the schematic of simulation setup. 
	A flow with the domain size $ L=0.004m $ and $ d=0.001m $ is driven by 
	a inlet and outlet pressure of $0.2 Pa$ and $0.1 Pa$, respectively, 
	for a constant pressure difference of $ \mathtt{\Delta} P= 0.1 Pa $. 
	In this simple unidirectional flow, 
	the bidirectional buffers is only used for 
	particle generation at the inlet, and deletion at the outlet. 
	The dynamic viscosity is set as 
	$ \eta= \sqrt{\rho d^{3} \mathtt{\Delta} P / 8L Re} $ with the fluid density 
	$ \rho=1000kg/m^{3} $ and the Reynolds number $ Re=50$. The fluid and 
	solid-wall particles are initialized on the Cartesian Lattice with a uniform particle 
	spacing of $ dp=d/50 $. 
	Note that, the artificial sound speed $ c_{0} $ 
	is chosen here with a large value as $ 100\mathbf{v}_{x}^{max} $, 
	with $ \mathbf{v}_{x}^{max}=d^{2}\mathtt{\Delta} P/8 \eta L $, 
	to ensure that the fast startup response is captured accurately 
	by the weakly compressible model.

\begin{figure}[htbp]
    	\centering     
    	\includegraphics[width=0.8\textwidth]{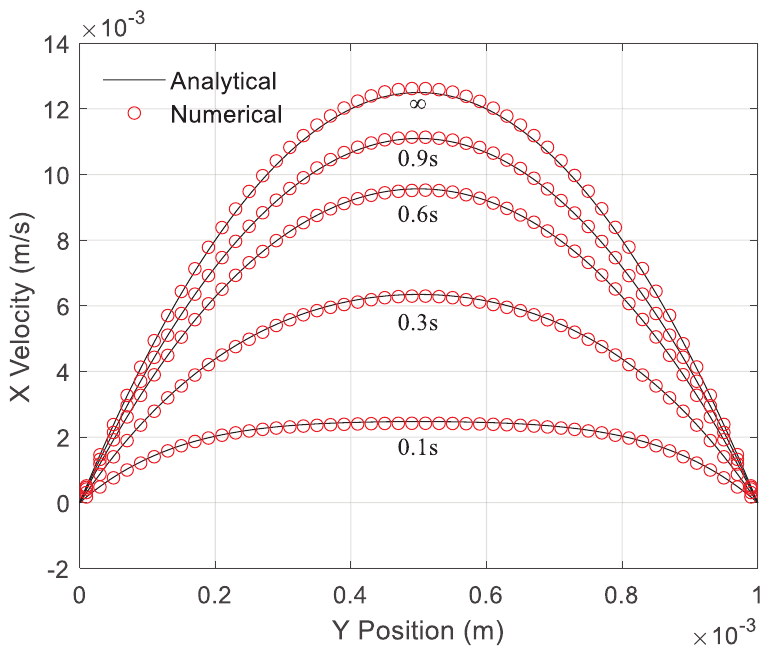}
    	\caption{Comparison of numerical and analytical velocity profiles at 5 
    		instants of the Poiseuille flow. The errors at 5 instants, 
    		from $ t=0.1s $ to $ \infty $, are $ 1.81\% $, $ 0.95\% $, $ 0.67\% $, 
    		$ 0.86\% $ and $ 1.22\% $, respectively.}
    	\label{Poiseuille_flow_velocity_comparison}
    \end{figure}

	In Figure \ref{Poiseuille_flow_velocity_comparison}, velocity profiles at 5 
	instants are extracted and compared with their corresponding 
	analytical solutions.
	Throughout the flow development process, the obtained velocity profiles 
	consistently exhibit a parabolic shape that closely aligns with the analytical solution.
	The maximum velocity occurs at the centerline, and the velocity approaches 
	zero near the plates as the no-slip solid-wall boundary condition \cite{ADAMI2012gene} 
	is applied.   
	Following the work of Holmes et al. \cite{HOLMES2021Novel}, an accuracy measurement 
	is also carried out with the Root Mean Square Error Percentage (RMSEP) over 
	the whole velocity profile

	\begin{equation} \label{RMSEP}
		RMSEP =\sqrt{\dfrac{1}{N}\sum_{n=1}^N \left(\dfrac{\mathbf{v}_{x}(y_{n},t)
				-\mathbf{\widetilde{v}}_{x}(y_{n},t)}
			{\mathbf{v}_{x}(y_{n}, t)}\right)^{2}},
	\end{equation}
	where $ N $ is the number of measuring points at a given time $ t $, 
	$ \mathbf{v}_{x}(y_{n}, t) $ and $\mathbf{\widetilde{v}}_{x}(y_{n},t) $ 
	the analytical and numerical velocities, respectively.
    The largest error of $ 1.81\% $ occurs at the start-up time $ t = 0.1s $, 
    while errors at other instants are around $ 1\% $, implying good agreement 
    with the analytical solution.

	\subsection{Startup of Hagen-Poiseuille flow}
	\label{Hagen_Poiseuille_flow} 
	To further verify the effectiveness of the proposed dynamical pressure boundary 
	condition in 3-D case, the startup of Hagen-Poiseuille flow, 
	which is the direct analogue to the Poiseuille flow in a circular pipe, is studied here, 
	while all other physical and geometrical parameters remain unchanged. 
	The analytical velocity evolution $\mathbf{v}_{x}(r,t)$ is given as

	\begin{equation} \label{Hagen_Poiseuille_profile}
		\mathbf{v}_{x}(r,t) = \frac{\mathtt{\Delta} P}{4\eta L}(R^{2}-r^{2})
		-\sum_{n=1}^\infty \dfrac{\mathtt{\Delta} PR^{2}}{\eta L\alpha_{n}^{2}}
		\dfrac{J_{2}(\alpha_{n})}{J_{1}^{2}(\alpha_{n})}
		J_{0}(\dfrac{r \alpha_{n}}{R})
		\exp(-\dfrac{\eta \alpha_{n}^{2}}{\rho R^{2}}t), 
	\end{equation}
    where $ r \in (0,R)$ with $ R=d/2 $ denoting the pipe radius. Here 
    $ J_{n} $ ($ n=0,1,2 $) are the Bessel functions of the first kind, 
    and $ \alpha_{n}$ ($ n=1,2,... $) are the roots of $ J_{0} $.
	The dynamic viscosity $ \eta= \sqrt{\rho R^{2} 
		\mathtt{\Delta} P / 4Re} $ 
	and, similarly, a large artificial sound speed 
	$ c_{0}=100\mathbf{v}_{x}^{max} $ with $ \mathbf{v}_{x}^{max}=R^{2}
	\mathtt{\Delta} P/4 \eta L $ are applied.
	
\begin{figure}[htb]
		\centering     
		\includegraphics[width=0.8\textwidth]{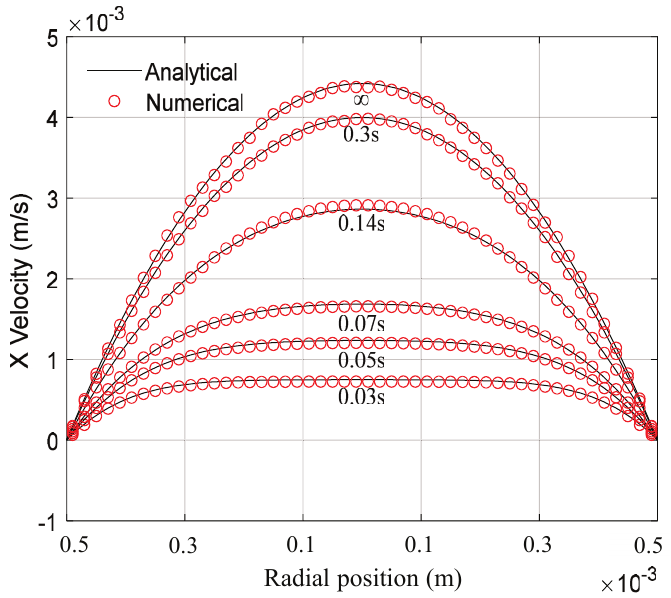}
		\caption{Comparison of numerical and analytical velocity profiles at 6 
			instants of the Hagen-Poiseuille flow. The errors at 6 instants, 
			from $ t=0.03s $ to $ \infty $, are $ 2.97\% $, $ 1.88\% $, $ 1.61\% $, 
			$ 1.5\% $, $ 0.74\% $ and $ 0.89\% $, respectively.}
		\label{Hagen_Poiseuille_flow_velocity_comparison}
	\end{figure}

	Figure \ref{Hagen_Poiseuille_flow_velocity_comparison} illustrates a comparison 
	between numerical and analytical velocity profiles at 6 instants. 
	The fluid velocity varies along the radial direction, with the maximum velocity 
	occurring at the centerline and progressively decreasing towards the pipe's wall, 
	which 
	also follows the parabolic velocity profile. 
	Furthermore, similar with the 2-D Poiseuille flow, the largest RMSEP, approximately 
	$2.97\%$, occurs at the start-up time $t = 0.03s$, and the error gradually diminishes 
	to within $1\%$ over time. 
	This error range also demonstrates a good agreement with the analytical solution 
	in the 3-D case.

\begin{figure}[htb]
		\centering     
		\includegraphics[width=0.8\textwidth]{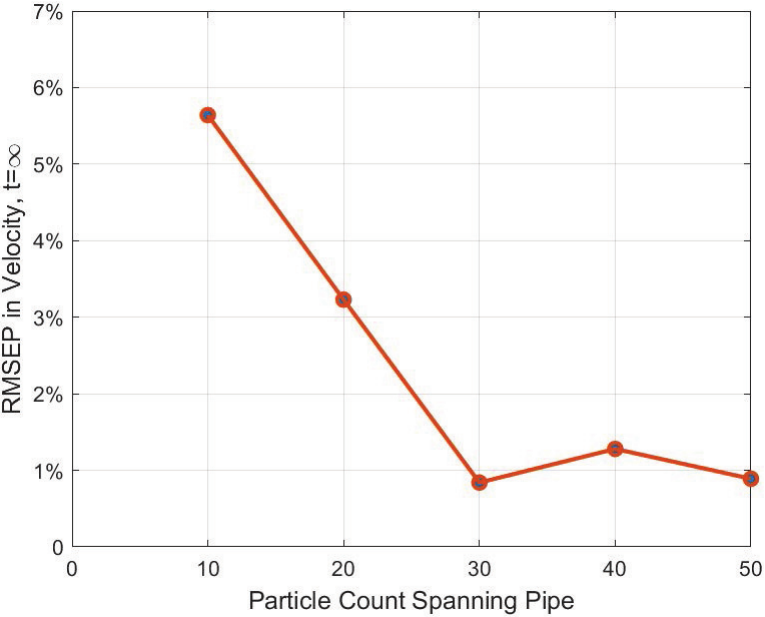}
		\caption{Plot of RMSEP for the whole velocity profile at time $ t=\infty $ for 
			Hagen-Poiseuille flow at different particle resolutions.}
		\label{Convergence_study}
	\end{figure}

	Referring to the convergence study on the Hagen-Poiseuille flow conducted by 
	Holmes et al. \cite{HOLMES2021Novel}, the accuracy and convergence of the present 
	dynamical pressure boundary condition are also investigated with varying spatial 
	resolution herein.
	Figure \ref{Convergence_study} illustrates the RMSEP across the entire velocity 
	profile at time $ t=\infty $ for various particle resolutions.
	As the number of particles spanning the circular pipe, i.e. $ 2R/dp $, exceeds 30, 
	the errors reach a saturation regime and are sufficiently small 
	(around $ 1\% $) compared to the analytical solution. 
	This level of accuracy and convergence closely aligns with that found in the 
	study by Holmes et al. \cite{HOLMES2021Novel}.

	\subsection{PIVO and VIPO channel flows}
	\label{Mixed_Poiseuille_flow}
	As the setups of PIVO and VIPO are commonly ultilized 
	in practical simulations, both PIVO and VIPO channel flows 
	are considered here.
	Based on the Poiseuille flow setup in Sec.
	\ref{Poiseuille_flow}, the pressure in-/outflow boundary
	conditions are replaced by the velocity boundary condition 
	in turn, and the prescribed velocity profile is analytically
	determined by Eq. (\ref{Poiseuille_profile}) at time 
	$ t=\infty $.
	
\begin{figure}[htbp]
		\centering     
		\includegraphics[width=0.9\textwidth]{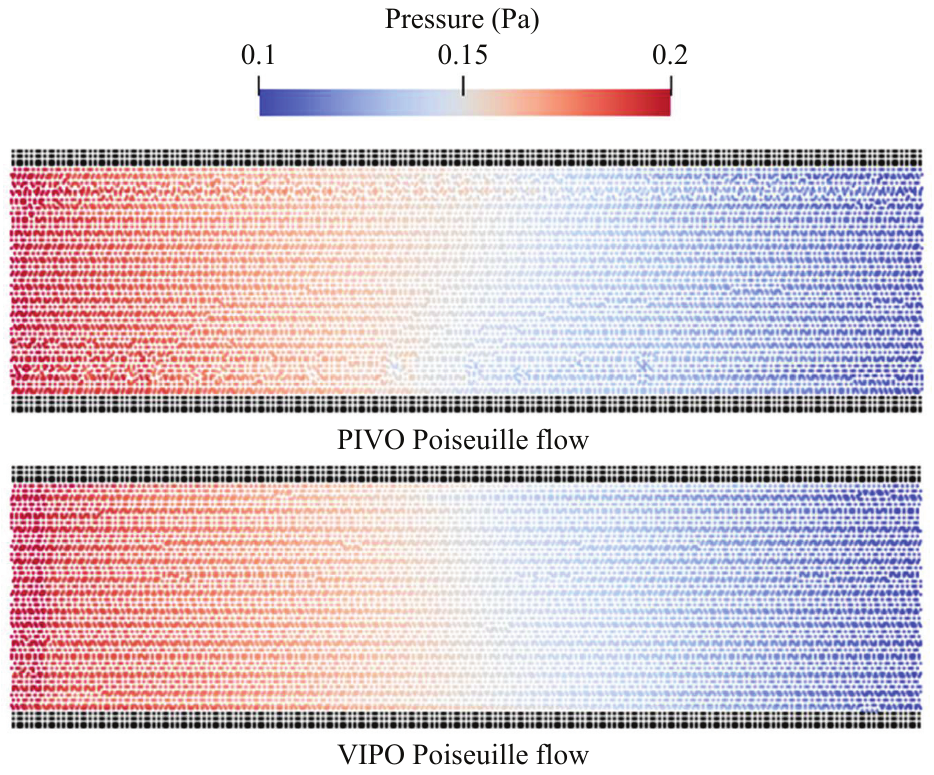}
		\caption{Pressure contours of PIVO (top panel) and VIPO (bottom panel) 
			channel flows. 
			For both cases, the left and right boundaries are the inlet and outlet, respectively.}
		\label{pressure_contour}
	\end{figure}

\begin{figure}[htbp]
		\centering     
		{
			\begin{minipage}{0.45\linewidth}
				\includegraphics[width=1\textwidth]{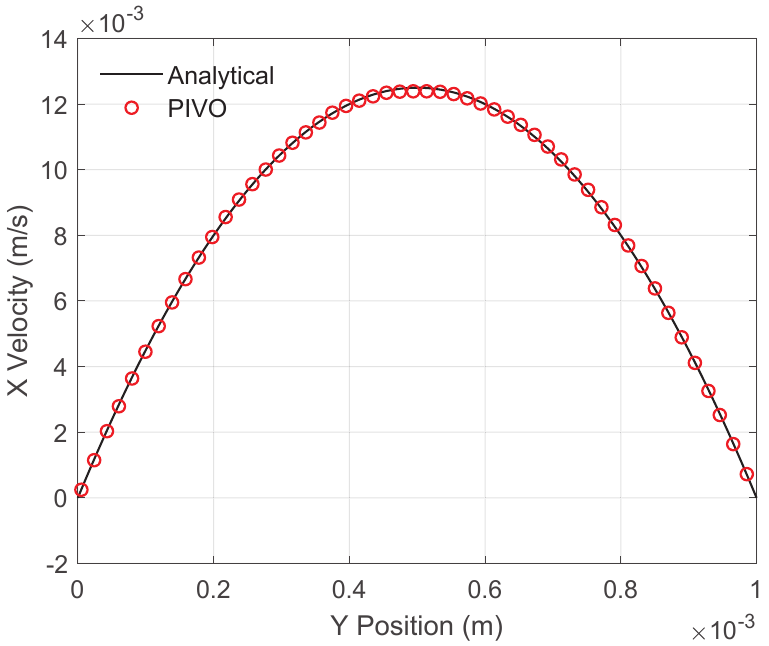}
			\end{minipage}
			\begin{minipage}{0.45\linewidth}
				\includegraphics[width=1\textwidth]{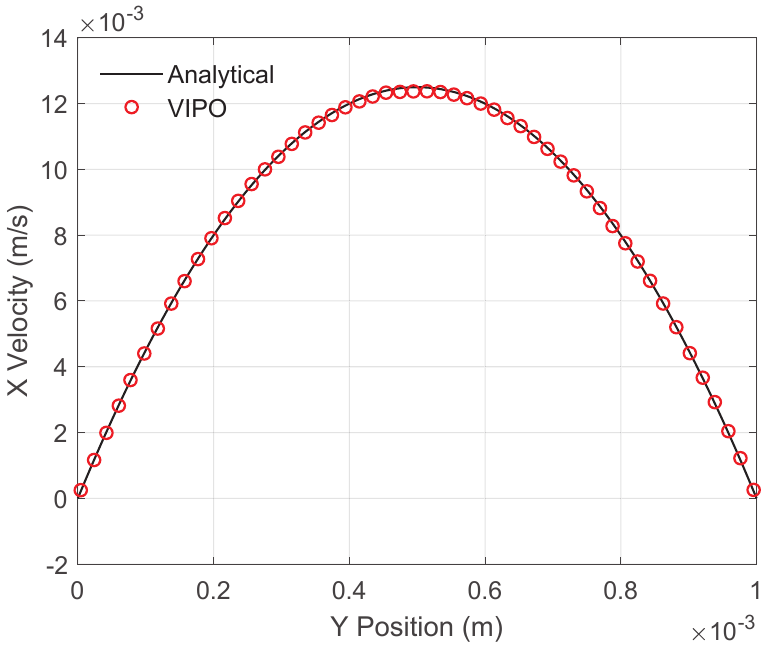}
			\end{minipage}
		}	
		\caption{Comparison of numerical and analytical velocity profiles of PIVO 
			(left panel) and VIPO (right panel) channel flows.
			The errors for PIVO and VIPO channel flows are $ 1.07\% $ and $ 2.32\% $, respectively.}
		\label{PIVO_VIPO_velocity}
	\end{figure}

	According to the setup parameters, the pressure difference between the 
	inlet and outlet should remain constant at $ \mathtt{\Delta} P= 0.1 Pa $. 
	In detail, in PIVO flow, the predicted 
	boundary pressure at the velocity outlet should be kept at $ 0.1 Pa $, 
	while in VIPO flow, 
	the predicted boundary pressure at the velocity intlet should be kept at $ 0.2 Pa $.
	Figure \ref{pressure_contour} gives the pressure distributions for both flows 
	and it is observed that the results are aligning well with the theoretical expectations.
	Figure \ref{PIVO_VIPO_velocity} 
	compares the computed velocity profiles of both PIVO and VIPO channel 
	flows with the corresponding analytical solution 
	and obtained small errors of $1.07 \%$ and $2.32 \%$ 
	respectively, indicating again good agreement with the analytical solution.


	\subsection{Pulsatile channel flow}
	\label{Pulsatile_flow}
\begin{figure}[htbp] 
    	\begin{adjustbox}
    		{addcode={\begin{minipage}{\width}}{\caption{The pressure gradient as a cosine function 
    						(upper panel) in the Pulsatile flow, and the comparison (lower panel) of 
    						numerical and analytical velocity profiles at 13 indicative instants. 
    						The horizontal grid dimension in the lower panel spans a velocity range 
    						from $ 0 $ to 5.0E-3 $m/s $. The errors at 13 instants, from 
    						$ t=2\pi $ to $ 4\pi $, are $ 1.45\% $, $ 2.24\% $, 
    						$ 2.65\% $, $ 3.98\% $, $ 3.11\% $, $ 2.56\% $, $ 2.83\% $, 
    						$ 2.53\% $, $ 2.61\% $, $ 3.22\% $, $ 2.55\% $, $ 2.54\% $ and $ 1.39\% $,  respectively.}\label{fig:Pulsatile_flow_velocity_comparison}\end{minipage}},
    					rotate=90,center}   		
    		\includegraphics[width=0.8\textwidth,angle=-90]{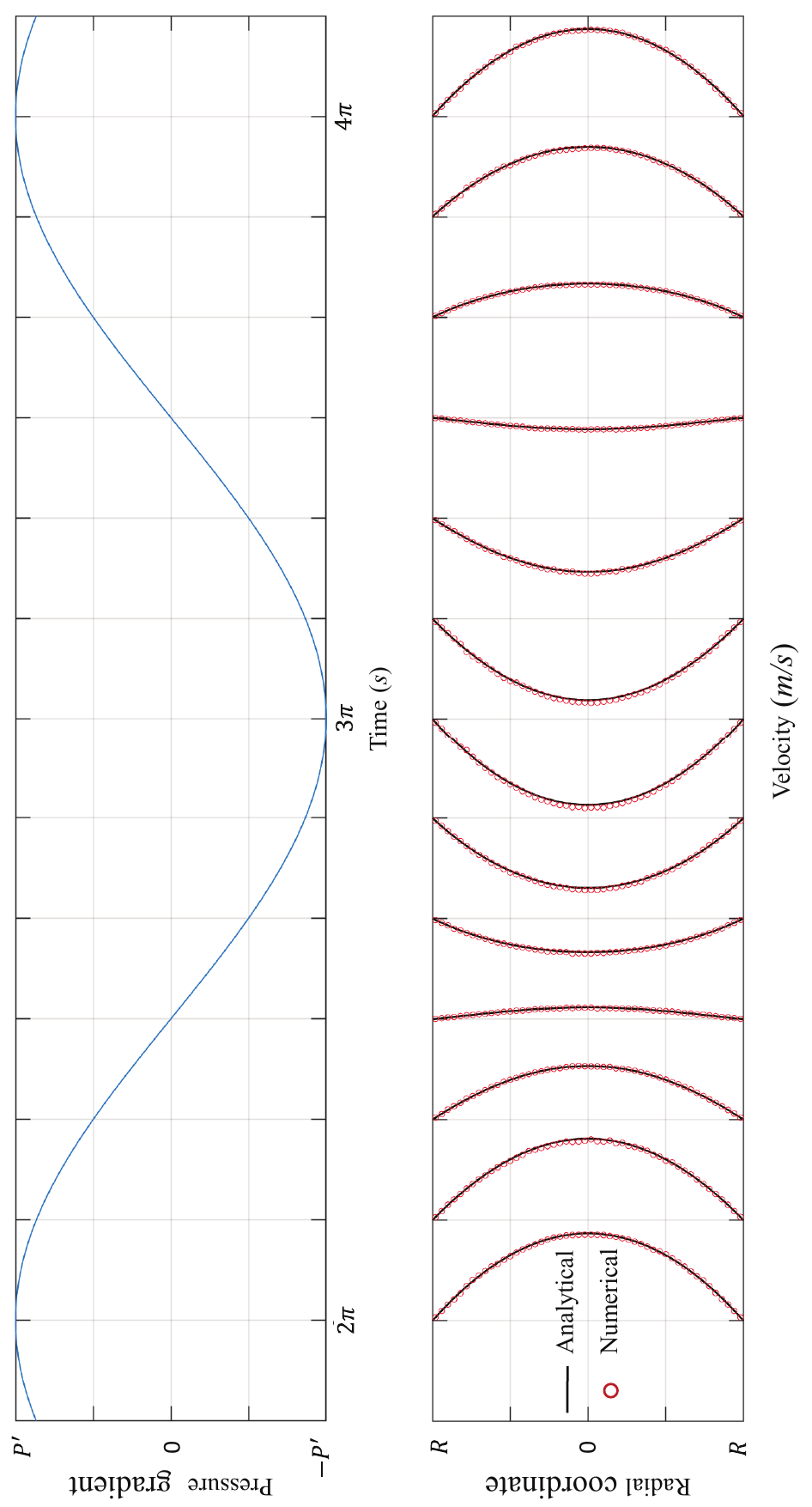}   		
    	\end{adjustbox} 
    \end{figure}

	The dual functionality of the bidirectional in-/outflow buffer, serving for both 
	particle generation and deletion, is specifically validated here through 
	a pulsatile channel flow. 
	To be more specific, building upon the Hagen-Poiseuille flow model discussed in 
	Subsection \ref{Hagen_Poiseuille_flow}, its constant pressure gradient is modified 
	to follow a cosine function over time, i.e., $P^{'}=\mathtt{\Delta} P/L=0.1\cos(t)/0.004 
	=25\cos(t)$, as illustrated in the upper panel of 
	Figure \ref{fig:Pulsatile_flow_velocity_comparison}. 
	The pulsatile flow profile was first derived by John R. Womersley for studying the blood 
	flow in arteries \cite{WOMERSLEY1955met}, and the analytical solution is given as
	\begin{equation} \label{Pulsatile_flow_profile}
		\mathbf{v}_{x}(r,t) = \mathbf{Re}\bigg\{\sum_{n=0}^N \dfrac{i P_{n}^{'}}
		{\rho n \omega}\bigg[1-\dfrac{J_{0}(\alpha n^{\frac{1}{2}}i^{\frac{3}{2}}
			\frac{r}{R})}{J_{0}(\alpha n^{\frac{1}{2}}i^{\frac{3}{2}})}\bigg]e^
		{in\omega t}\bigg\}, 
	\end{equation}
    where $ r \in (0,R)$, $ \alpha=R\sqrt{\omega \rho/\eta} $ the dimensionless 
    Womersley number, $ \omega $ the angular frequency of the first harmonic of 
    a Fourier series of an oscillatory pressure gradient, $ P_{n}^{'} $ the pressure 
    gradient magnitude for the frequency $ n\omega $, $ J_{0}(\cdot) $ the Bessel 
    function of first kind and order zero, $ i $ the imaginary number, and 
    $ \mathbf{Re}\left\{\cdot\right\}$ the real part of a complex number.
    
    It's well-known that the shape of the pulsatile flow profile changes depending 
    on the Womersley number $ \alpha $. For 
    example, a flattened velocity profile for $ \alpha \gtrsim 2$ and a parabolic 
    profile for $ \alpha \lesssim 2$.
    In the present pulsatile flow with $ \alpha = 0.8409$, suggesting viscous forces dominate 
    the flow and resulting in a quasi-static pulse with a parabolic profile. 
    Moreover, the time evolution of the velocity profile exhibits periodicity in both 
    velocity magnitude and 
    direction, which is consistent with the cosine-varying 
    pressure gradient.
    In the lower panel of Figure \ref{fig:Pulsatile_flow_velocity_comparison}, 
    during one period from $ t = 2\pi  $ to $ 4\pi $, the obtained velocity profiles 
    extracted at 13 instants  accurately follow the periodically evolving 
    parabolic shape.
    In more detail, the parabolic velocity profile maintains a forward flow direction 
    during the first quarter, but with a decreasing magnitude over time. 
    At $ t = 2.5\pi $, the flow direction reverses, and the flow gradually accelerates 
    until it reaches the maximum velocity at $ t = 3\pi$.
    In the latter half of the period, 
    this flow behavior will repeat, but with the opposite evolution 
    of flow direction.
  
    In the accuracy measurement using RMSEP, the numerical results also exhibit a good 
    fit with their corresponding analytical solutions, with an overall error of 
    approximately $ 3\% $ for all 13 instants. For the objectives of this 
    study, this level of accuracy is considered acceptable.

	\subsection{Generic carotid and aorta flows}
    \label{Aorta_flow}
\begin{figure}[htbp]
	\centering     
	\includegraphics[width=\textwidth]{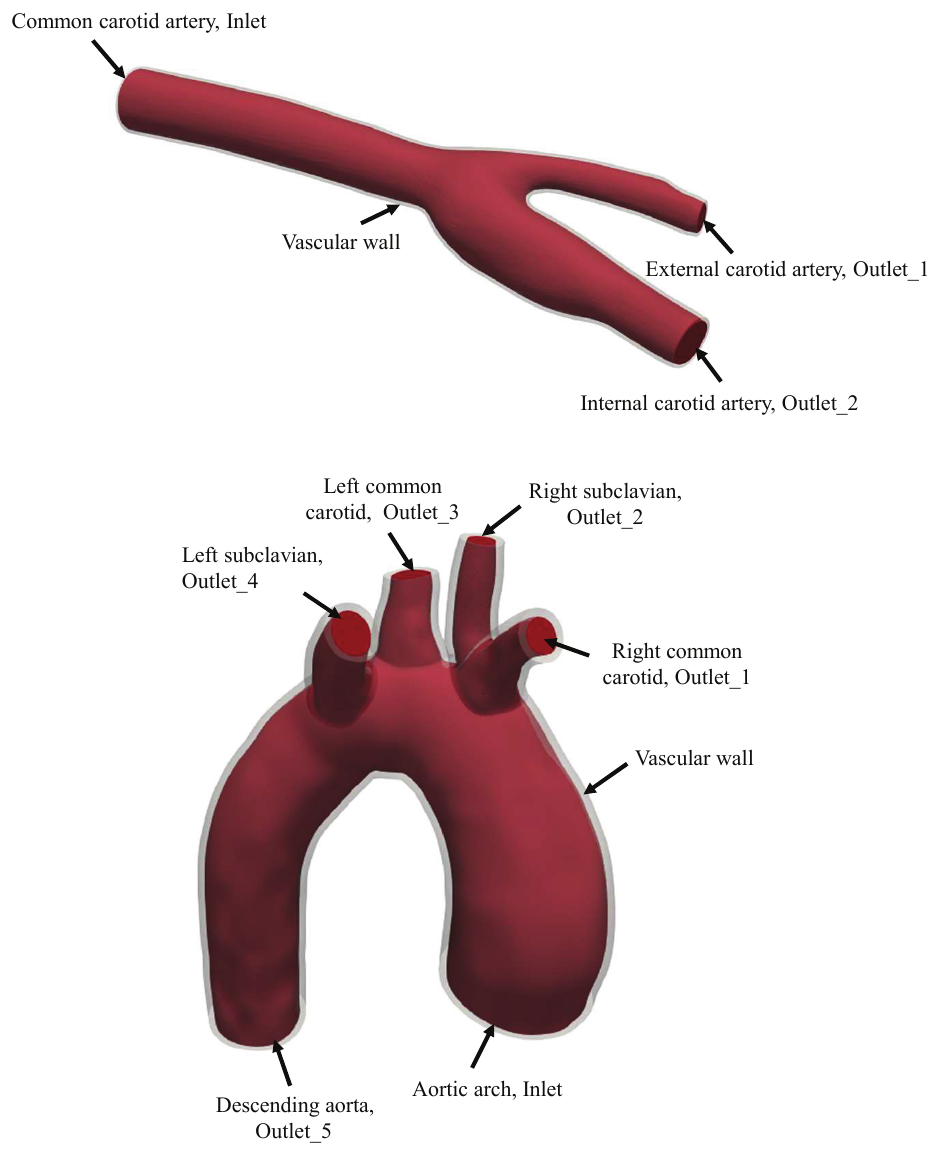}
	\caption{Schematic of the generic carotid and aorta flows.}
	\label{vessel_schematic}
    \end{figure}

    To investigate the applicability and versatility of the 
    present dynamical pressure boundary condition for hemodynamic simulations, 
    flow simulations 
    are preliminarily conducted on carotid and aorta structures, 
    as shown in Figure \ref{vessel_schematic}. 
    The blood is modeled as Newtonian fluid with density of 1060 $ kg/m^{3} $ \cite{Katia2005Multi} 
    and 
    viscosity of 0.00355 $ Pa\cdot s $ \cite{Gallo2012Vivo}, and the vascular wall is simplified 
    as a rigid body. At inlet, a velocity inflow boundary condition with a plug profile 
    is implemented, 
    while the pressure boundary condition is enforced across all outlets. Note that, 
    to increase the computational efficiency, 
    the standard choice of artificial sound speed  $ c_{0}=10 U_{max} $ 
    is applied \cite{Morris1997Modeling}.

    For the carotid flow, considering the pulsatile nature of blood circulation, the inflow 
    velocity is defined as a waveform function of time, with the flow direction aligned 
    along the normal vector of the inlet boundary surface
    \begin{equation} \label{piecewise_function}
	\mathbf{v}_{inlet}(t)=\left\{
	\begin{array}{cc} 
		0.5sin[4\pi(t+0.0160236)] & 0.5n<t \leq 0.5n+0.218 \\[3mm]
		0.1 & 0.5n+0.218<t \leq 0.5(n+1)\\
	\end{array}\right.,
    \end{equation} 
    where $ n=0, 1, 2... $ Furthermore, because the blood pressure 
    for a healthy person is around $ 120 mmHg/80mmHg $ (systolic/diastolic), 
    the constant average blood pressure $ 100 mmHg $ is imposed at both outlets.

\begin{figure}[htbp]
	\centering     
	\includegraphics[width=0.53\textwidth]{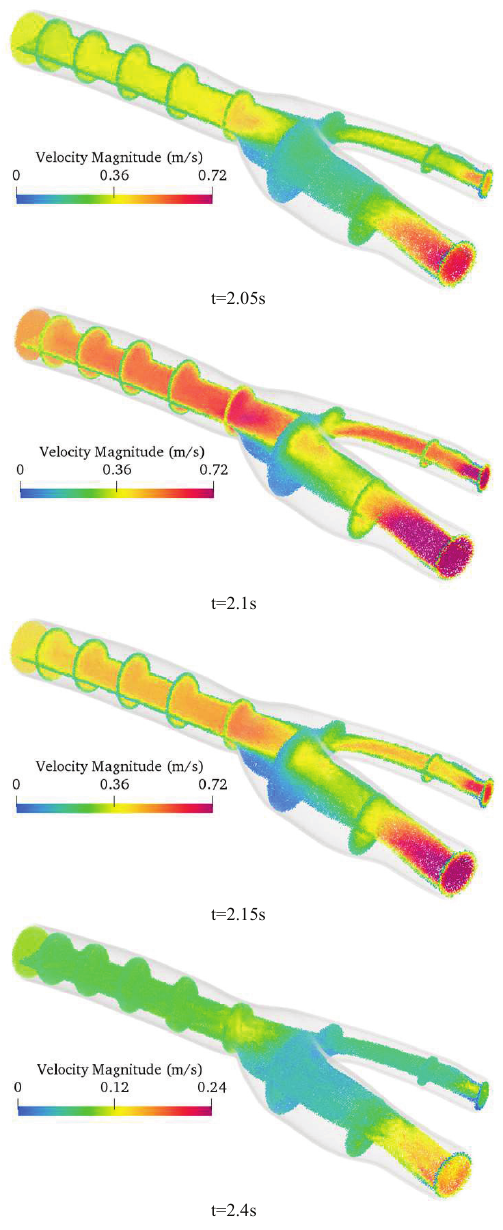}
	\caption{Velocity contours of carotid flow at 4 time instants of the 5th cycle. 
	From top to bottom panels, the maximum velocities are around 0.72 $ m/s $, 1.0 $ m/s $, 0.86 $ m/s $ 
	and 0.24 $ m/s $, respectively.}
	\label{velocity_contour}
    \end{figure}

\begin{figure}[htbp]
	\centering     
	\includegraphics[width=0.8\textwidth]{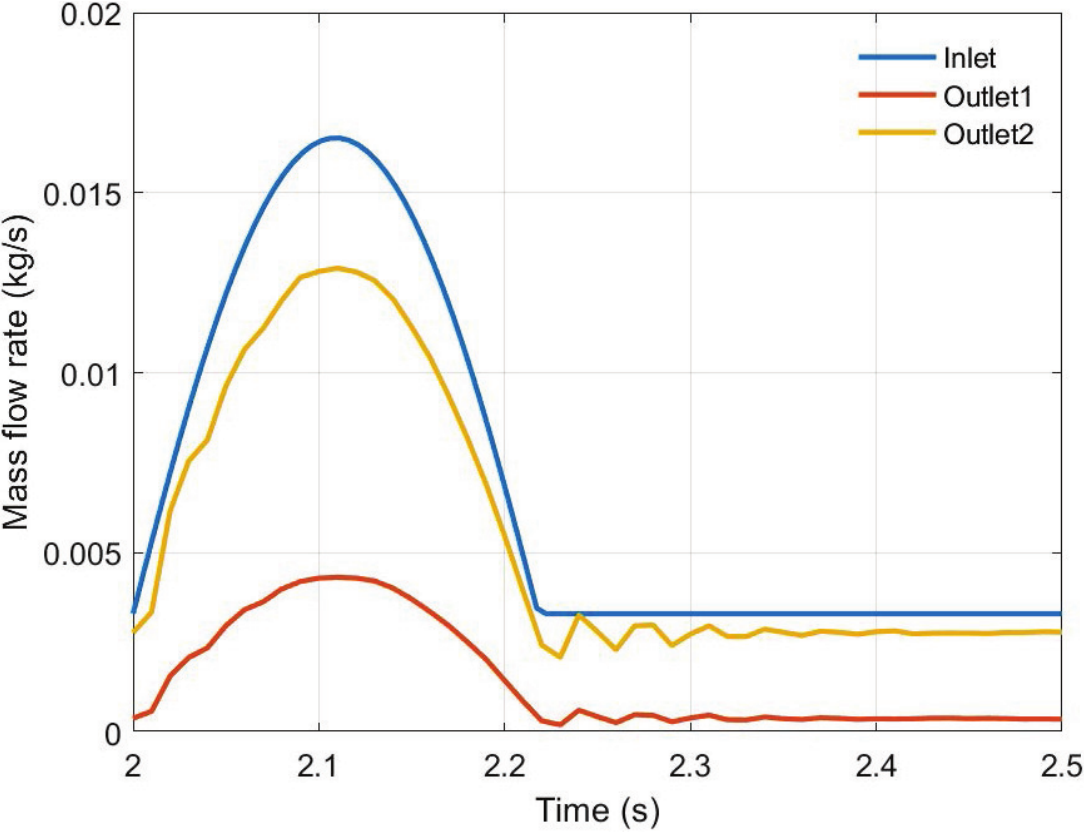}
	\caption{Time histories of mass in-/outflow rates in carotid flow. 
		Both mass outflow rates are numerically measured, while the mass
		inflow rate is determined analytically.}
	\label{mass_flow_rate}
    \end{figure}

\begin{figure}[htbp] 
    	\begin{adjustbox}
    		{addcode={\begin{minipage}{\width}}{\caption{The time history 
    						of the inflow velocity and all outlet relative pressure.}
    					\label{fig:Aorta_relative_pressure}\end{minipage}},rotate=90,center}   		
    		\includegraphics[width=1.45\textwidth,angle=0]{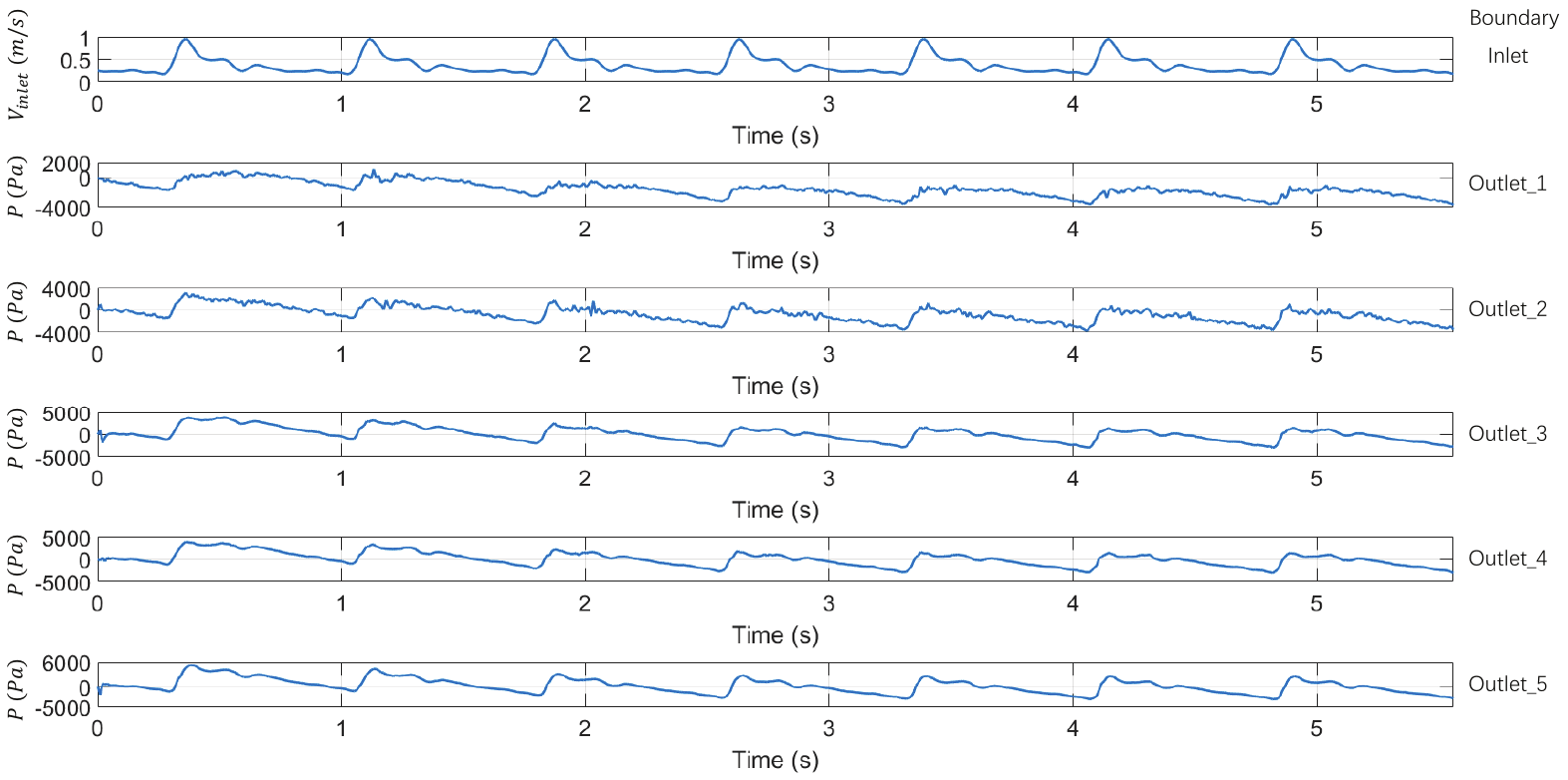}   		
    	\end{adjustbox} 
    \end{figure}

    Figure \ref{velocity_contour} illustrates the velocity contours at 4 time instants 
    during the 5th cycle. The blood flows into the bifurcation artery from the common 
    carotid artery, and then the flow splits into the internal and external carotid 
    arteries. Due to the geometric asymmetry of the blood vessel, the velocity magnitude 
    and distribution in the internal and external carotid arteries are notably distinct. 
    Moreover, the time-dependent pulsatile inflow also induces corresponding changes in 
    the entire velocity field over time.
    To demonstate the accuracy and reliability of the present simulation, especially for 
    the pulsatile characteristic, the mass flow rate during the 5th cycle is chosen to 
    be analyzed.
    In Figure \ref{mass_flow_rate}, the time histories of both mass outflow rates 
    correspond well to that of pulsatile mass inflow rate. 
    Note that, due to the weakly compressible assumption in WCSPH method, 
    mass flow rate fluctuates slightly in agreement with Ref. \cite{HOLMES2021Novel}. 
    For the aorta flow, the waveform of inflow velocity in Figure \ref{fig:Aorta_relative_pressure} 
    is approximated as Fourier series in term of time \cite{Taib2019blood}, 
    which can mimic the continuous cardiac cycle of the waveform 
    \begin{equation} \label{Fourier_Series}
    	\mathbf{v}_{inlet}(t)= a_{0}+\sum_{n=1}^8 a_{n}cos(nt\omega)+
    	\sum_{n=1}^8 b_{n}sin(nt\omega),
    \end{equation} 
    where the empirical variables for the Fourier series are listed in 
    Table \ref{empirical_variables}.   
    \begin{table}[htbp]
    	\centering
    	\caption{The empirical variables for the Fourier series \cite{Taib2019blood}}
    	\resizebox{0.6\textwidth}{!}{
    		\begin{tabular}{cccc}
    			\toprule  
    			Parameter  & \multicolumn{1}{c}{Value} & \multicolumn{1}{c}{Parameter} 
    			& \multicolumn{1}{c}{Value} \\
    			\midrule 
    			$ a_{0} $ & 0.3782 & $ \omega $ & 8.302 \\
    			$ a_{1} $ & -0.1812 & $ b_{1} $ & -0.07725 \\
    			$ a_{2} $ & 0.1276 & $ b_{2} $ & 0.01466 \\
    			$ a_{3} $ & -0.08981 & $ b_{3} $ & 0.04295 \\
    			$ a_{4} $ & 0.04347 & $ b_{4} $ & -0.06679 \\
    			$ a_{5} $ & -0.05412 & $ b_{5} $ & 0.05679 \\
    			$ a_{6} $ & 0.02642 & $ b_{6} $ & -0.01878 \\
    			$ a_{7} $ & 0.008946 & $ b_{7} $ & 0.01869 \\
    			$ a_{8} $ & -0.009005 & $ b_{7} $ & -0.01888 \\
    			\bottomrule     			    
    	\end{tabular}}
    	\label{empirical_variables} 
    \end{table}

    Different from the constant outlet pressure in the carotid flow, the outlet 
    pressure $ P(t) $ in the aorta flow is time-dependent and obtained by 
    a 3-element (RCR) Windkessel model \cite{Catanho2012Model}
    \begin{equation} \label{windkessel}
    	(1+\dfrac{R_{p}}{R_{d}})Q(t)+CR_{p}\frac{dQ(t)}{dt}=\dfrac{P(t)}
    	{R_{d}}+C\dfrac{dP(t)}{dt},
    \end{equation} 
    where $ Q(t) $ is the blood flow volume,  $ C $ the arterial compliance, 
    $ R_{p} $ the characteristic resistance, and $ R_{d} $ the peripheral resistance. 
    The values of Windkessel parameters \cite{Kim2007coupling, Sudharsan2018effect} 
    used for the various daughter vessels are listed in Table \ref{Windkessel_parameters}.

    \begin{table}[htbp]
    	\centering
    	\caption{Parameters for the Windkessel outlet boundary 
    		conditions \cite{Kim2007coupling, Sudharsan2018effect}}
    	\resizebox{\textwidth}{!}{
    		\begin{tabular}{cccc}
    			\toprule  
    			Vessel  & \multicolumn{1}{c}{\textbf{$ R_{p}(dynes\: s/cm^{5}) $}} 
    			& \multicolumn{1}{c}{\textbf{$ C(cm^{5}/dynes) $}} & \multicolumn{1}{c}
    			{\textbf{$ R_{d}(dynes\: s/cm^{5})$}} \\
    			\midrule 
    			Right common carotid &  1180     &  7.70E-5  &  18400      \\
    			Right subclavian&     1040      &  8.74E-5        &     16300     \\
    			Left common carotid&   1180     & 7.70E-5 &      18400 \\
    			Left subclavian&      970     & 9.34E-5   &     15200     \\
    			Descending aorta&    188      &    4.82E-4      &     2950      \\
    			\bottomrule     			    
    	\end{tabular}}
    	\label{Windkessel_parameters} 
    \end{table}

\begin{figure}[htbp]
    	\centering     
    	\includegraphics[width=0.6\textwidth]{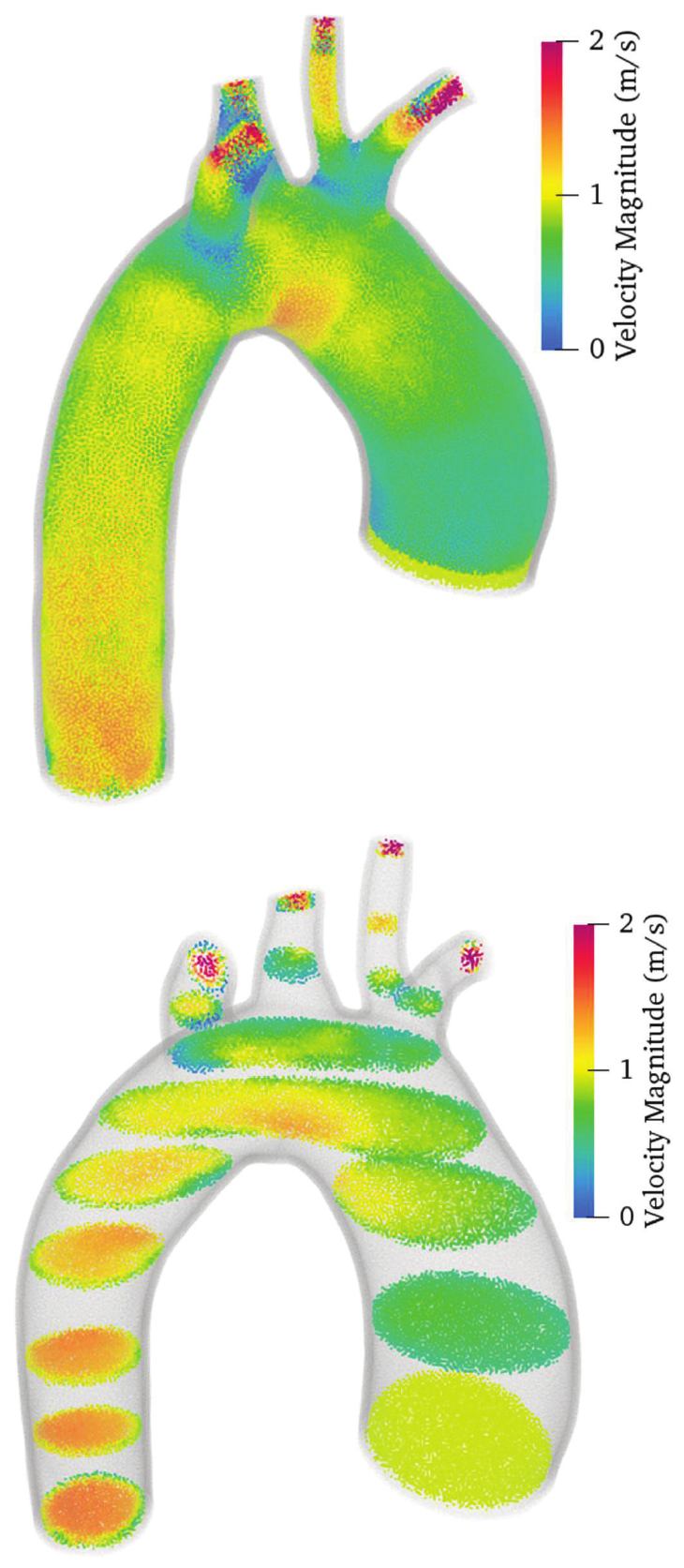}
    	\caption{Velocity contour and slices of aorta flow during the peak systole. 
    		Time instant $ t=4.9s $.}
    	\label{Aorta_velocity_contour}
    \end{figure}

    Corresponding to the time history of inflow velocity over 7 cycles, 
    Figure \ref{fig:Aorta_relative_pressure} shows the relative pressure at all outlets. 
    Same as both analytical and numerical solutions for the blood pressure in 
    Ref. \cite{Catanho2012Model}, 
    all the waveforms of the present outlet relative pressure exhibit a folding 
    line shape, which is consistent with theoretical expectations.
    Note that, compared to the practical physiological scenario, the present 
    inflow velocity, the Windkessel parameters of all daughter vessels, and 
    the aorta geometry don't strictly match each other. Moreover, 
    the corresponding physiological pressure data for the present
    aorta model are lacking. Hence, our primary 
    focus is on the accuracy of the blood pressure waveform rather than the 
    pressure magnitude.
    Figure \ref{Aorta_velocity_contour} also illustrates the velocity contour 
    and slices during the peak systole.
    In the present plug-inlet boundary condition, the velocity peak isn't 
    always at the center of the cross-section. Especially in the descending 
    aorta, due to the inertia, the flow inside the curved vessel is pushed
    towards the outer side of the arch, 
    which is in good agreement with the observations in Ref. \cite{Sudharsan2018effect}.

	\section{Conclusion}
	\label{Conclusion}
	In this paper, we develop a dynamical pressure boundary condition for the WCSPH 
	method, characterizing a bidirectional in-/outflow buffer to simulate 
	hemodynamic flows with complex geometric boundaries. 
	With the assumption of zeroth-order consistency in the SPH discretization, the 
	dynamical boundary pressure is imposed by the SPH approximation 
	of the pressure gradient on near-boundary particles. 
	The boundary pressure could be constant, 
	time-dependent, or dynamically determined by the Windkessel model 
	as in practical hemodynamic simulations.
	The bidirectional in-/outflow buffer is achieved 
	by relabelling buffer particles at each time step, 
	and is suitable for simulating both uni-/bidirectional flows, 
	especially those with mixed in-/outflow boundary condition.
	The numerical results in 2-D and 3-D benchmark cases exhibit very good agreement 
	with analytical solutions, demonstrating the accuracy and effectiveness of the 
	present method. 
	As the present method performs well in the generic carotid and aorta flow simulations, 
	and thanks to the effectiveness of SPH method for FSI problems, 
	our future focus will center on exploring cardiovascular hemodynamics
	and other relevant bio-engineering applications 
	within a unified meshfree computational framework.
	
	
	
	\bibliographystyle{elsarticle-num} 
	\bibliography{reference_literature}
	
	
	
	
	%%\bibitem{}
	
	%%\end{thebibliography}
\end{document}